# Decoherence in elastic and polaronic transport via discrete quantum states

Kamil Walczak

Institute of Physics, Adam Mickiewicz University
Umultowska 85, 61-614 Poznań, Poland

Here we study the effect of decoherence on elastic and polaronic transport via discrete quantum states. The calculations are performed with the help of nonperturbative computational scheme, based on the Green's function theory within the framework of polaron transformation (GFT-PT), where the many-body electron-phonon interaction problem is mapped exactly into a single-electron multi-channel scattering problem. In particular, the influence of dephasing and relaxation processes on the shape of the electrical current and shot noise curves is discussed in detail under the linear and nonlinear transport conditions.

## 1. Introduction

Molecular junctions are devices composed of individual molecules (or molecular layers) connected to two (or more) metallic electrodes, operating under the influence of a bias voltage. Good candidates for playing the role of molecular bridges are organic molecules [1,2] due to: (i) their delocalized pi-type orbitals suitable for transport (conduction channels between reservoirs of charge carriers), (ii) self-assembly features (helpful in fabrication), (iii) possibility of theoretically inexhaustible structural modifications of the molecules, and (iv) potential to become relatively cheap (important from the economical point of view). Transport properties at molecular scale strongly depend on some quantum phenomena, such as: (i) quantum tunneling, (ii) quantization of molecular energy levels, and (iii) discreteness of electron charge. Since molecules involved into the conduction process at finite temperatures can be thermally activated to vibrations (phonon modes are excited), their transport properties should be also affected by vibronic coupling.

Moreover, inelastic electron tunneling spectroscopy (IETS) is a powerful experimental tool for identifying and characterizing molecular species within the conduction region [3-19]. Standard ac modulation techniques, along with two lock-in amplifiers, are utilized to measure current-voltage ($I-V$) characteristics as well as the first and second harmonic signals (proportional to $dI/dV$ and $d^2I/dV^2$, respectively). This method provides information on the strength of the vibronic coupling between the charge carriers and nuclear motions of the molecules. The IETS experiment can also be helpful in identifying the geometrical structures of molecules and molecule-metal contacts, since junctions with different geometries disclose very different spectral profiles [18,19]. The measured spectra show well-resolved vibronic features corresponding to certain vibrational normal modes of the molecules. The IETS spectra are also very sensitive to the device working temperature and intramolecular conformational changes.

So far, inelastic transport was treated as a purely coherent transfer process [20-29]. However, the coherent tunneling cannot get very far at room temperatures, since the thermal disorder destroys the orbital delocalization leading to incoherent transport [30-34] and some of the experimental data may be interpreted in this sense [35-40]. Decoherence scattering rate can be introduced phenomenologically by means of the parameter $\tau_D$ that represents the scattering time of electrons with molecular vibrations (phonons). It should be noted that phase loss can also be induced by spin-flip scattering as well as spin-orbit processes. When the time $\tau_D$ is of order of magnitude comparable to the residence time of a tunneling electron on the molecule ($\sim fs$), the effect of dephasing cannot be ignored during the transport. Assuming $\tau_D \sim fs$, the corresponding energy parameter is $\Gamma_D = \hbar/\tau_D \sim 0.1$ eV. Obviously, decoherence is limited mainly by the extent of the molecular orbital. General tendency is as follows: the more localized the electronic wavefunction, the bigger the value of the $\Gamma_D$-parameter.

The main purpose of this work is to study the effect of dephasing and relaxation processes on elastic and especially on polaronic transport via discrete quantum states. The calculations are performed using nonperturbative computational scheme, based on Green's function theory within the framework of polaron transformation (GFT-PT). This method transforms the many-body electron-phonon interaction problem into a single-electron many-channel scattering problem [20-29]. Here we analyze the electrical current and shot noise (current fluctuations of purely quantum origin) in two response regimes: linear and nonlinear, respectively. Although noise measurements in molecular devices still remain a certain challenge, there are some theoretical suggestions in the literature retaining to that question [28,41-45].

## 2. Description of the model

Let us consider molecular quantum dot weakly connected to two electrodes through tunnel barriers, described by the Hamiltonian

$$H = \sum_{k \in \alpha} \varepsilon_k c_k^+ c_k + \sum_{k \in \alpha} \left( \gamma_k c_k^+ c_l + h.c. \right) + \varepsilon_l c_l^+ c_l + \Omega a^+ a - \lambda (a + a^+) c_l^+ c_l. \qquad (1)$$

The first term describes the left and right electrodes ($\alpha = L, R$) and the dephasing reservoir as well ($\alpha = D$), the second term describes the tunnel connection between the molecule and all the reservoirs, while the last three terms represent molecular part of the Hamiltonian, where one spin-degenerate electronic level is coupled to a single vibrational mode (primary mode). Here $\varepsilon_k$ and $\varepsilon_l$ are energies of electronic states in reservoirs and on the molecule, $\gamma_k$ is the strength of the molecule-reservoir connection, $\Omega$ is the phonon energy, $\lambda$ is the electron-phonon interaction parameter. Furthermore, $c_k$, $c_l$, $a$ and their adjoints are annihilation and creation operators for the electrons in reservoirs and on the molecular level, and for the primary phonon, respectively. In this work, we follow Büttiker's idea to include decoherence that the phase-breaking processes can be modeled by connecting the molecule with a fictitious electronic reservoir ($\alpha = D$) [46].

The problem we are facing now is to solve a many-body problem with phonon emission and absorption when the electron tunnels through the molecule. To carry out the calculations, we apply the so-called polaron transformation, where the electron states into the molecule are expanded onto the direct product states composed of single-electron states and $m$-phonon Fock states $|l,m\rangle = d_l^+ (a^+)^m |0\rangle / \sqrt{m!}$ (electron state $|l\rangle$ is accompanied by $m$ phonons, and $|0\rangle$ denotes the vacuum state). Similarly, the electron states in all the $\alpha$- reservoirs can be expanded onto the states $|k,m\rangle = c_k^+ (a^+)^m |0\rangle / \sqrt{m!}$. Such procedure enables us to map the many-body electron-phonon interaction problem into a multi-channel single-electron scattering problem, as shown in Fig.1 and discussed elsewhere [20-29]. After eliminating the degrees of freedom of the two electrodes and the dephasing reservoir, we can present the effective Hamiltonian of the reduced molecular system in the form

$$H_{eff} = \sum_{m,\alpha} \left( \varepsilon_l + m\Omega + \Sigma_\alpha^m \right) |l,m\rangle\langle l,m| - \sum_m \lambda\sqrt{m+1} \left( |l,m+1\rangle\langle l,m| + |l,m\rangle\langle l,m+1| \right). \quad (2)$$

Index $m$ determines the statistical probability to excite the phonon state $|m\rangle$ at finite temperature $\theta$, and therefore the accessibility of particular conduction channels is determined by a weight factor of the Boltzmann distribution function $P_m = [1 - \exp(-\beta\Omega)]\exp(-m\beta\Omega)$, where $\beta = 1/(k_B \theta)$ and $k_B$ is Boltzmann constant. In practice, the basis set is truncated to a finite number of possible excitations $m = m_{max}$ in the phonon modes because of the numerical efficiency. The size of the basis set strongly depends on: (i) phonon energy $\Omega$, the temperature of the system under investigation $\theta$, and the strength of the electron-phonon coupling constant $\lambda$. It should be noted that effective Hamiltonian (given by Eq.2) is analogous to tight-binding model with different site energies and sit-to-site hopping integrals (formed along the tunnel barriers, as can be seen in Fig.1). Obviously, each site is additionally connected with a dephasing reservoir in order to simulate the phase-breaking effect.

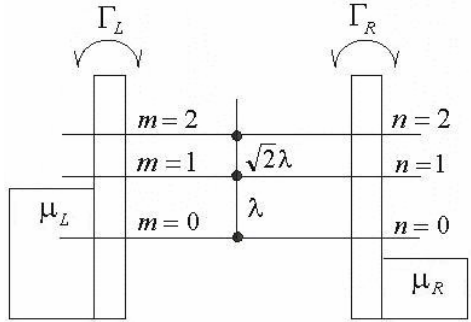

Figure 1: A schematic representation of inelastic scattering problem for the device composed of molecular quantum dot with single energy level connected to two metallic electrodes.

For simplicity, we adopt the wide-band approximation to treat both electrodes and the dephasing reservoir, where the self-energy and the so-called linewidth function are given through the relations $\Sigma_\alpha^m = -i\Gamma_\alpha^m/2$ and $\Gamma_\alpha^m = 2\pi |\gamma_k^m|^2 \rho_\alpha$, respectively. Here $\gamma_k^m \equiv \gamma_\alpha$ is the energy and voltage independent parameter (by assumption) related to the strength of the effective connection between the $m$ th channel and the $\alpha$-reservoir characterized by constant density of states $\rho_\alpha$. Both electrodes are also identified with their electrochemical potentials: $\mu_L = \varepsilon_F + \eta eV$, $\mu_R = \varepsilon_F - (1-\eta)eV$ which are related to the Fermi energy level $\varepsilon_F$ [36,37]. The voltage division factor $0 \leq \eta \leq 1$ describes how the electrostatic potential difference $V$ is divided between two contacts and can be related to the relative strength of the couplings with the electrodes $\eta = \gamma_R/(\gamma_R + \gamma_L)$.

Now we proceed to analyze the problem of electron transfer between two electrodes via discrete quantum state in the presence of phonons. An electron entering from the left hand side can suffer inelastic collisions by absorbing or emitting phonons before entering the right electrode. Such processes are presented graphically in Fig.1, where individual channels are indexed by the number of phonon quanta in the left ($m$) and right electrode ($n$), respectively. Each of the mentioned processes is described by its own transmission probability

$$T_{m,n} = T_{m,n}^{RL} + \frac{T_{m,n}^{RD} T_{m,n}^{DL}}{T_{m,n}^{RD} + T_{m,n}^{DL}}. \tag{3}$$

The physical meaning of Eq.3 is clear: the first term gives the coherent contribution to tunneling, while the second term gives incoherent component that originates from the loss of phase by the electrons scattered inside the molecule [36,37,46]. Particular coefficients

$$T_{m,n}^{\alpha'\alpha}(\varepsilon) = \Gamma_{\alpha'} \Gamma_{\alpha} |G_{m+1,n+1}(\varepsilon)|^2 \tag{4}$$

are given through the matrix element of the molecular Green function of size $m_{max} \times m_{max}$

$$G(\varepsilon) = [1\varepsilon - H_{eff}]^{-1}. \tag{5}$$

Here 1 stands for identity matrix, while $H_{eff}$ is the molecular Hamiltonian (Eq.2). The effect of the connections with the $\alpha$-reservoirs is fully described by specifying self-energy corrections $\Sigma_{\alpha}$. Having probabilities for all the possible transitions (Eq.3), we can define the total transmission function [22]

$$T(\varepsilon) = \sum_{m,n} P_m T_{m,n}. \tag{6}$$

The elastic contribution to the transmission can be obtained by the assumption $n = m$. It should be noted that at low voltages, the linear conductance ($G = I/V$) is directly proportional to transmission function (Eq.6) determined exactly at the Fermi energy $G = G_0 T(\varepsilon_F)$, where $G_0 = 2e^2/h \cong 77.5$ μS is the quantum of conductance.

The total current flowing through the junction can be expressed in terms of transmission probability of the individual transition $T_{m,n}$ (Eq.3) which connects incoming channel $m$ with outgoing channel $n$ [22]

$$I(V) = I_0 \int_{-\infty}^{+\infty} d\varepsilon \sum_{m,n} T_{m,n} \left[ P_m f_L^m (1 - f_R^n) - P_n f_R^n (1 - f_L^m) \right]. \tag{7}$$

Here $I_0 = 2e/h$, while $f_\alpha^m = [1 + \exp[\beta(\varepsilon + m\Omega - \mu_\alpha)]]^{-1}$ is the Fermi distribution function associated with the $\alpha$-electrode. Then the nonlinear conductance can be calculated as the derivative of the current with respect to voltage.

Shot noise is the time-dependent fluctuation of the electrical current due to the discreteness of the charge of the current carriers and can be computed as the Fourier transform of the current-current correlation function [47]. Limiting ourselves to the final results, the expression for the total spectral density of shot noise in the zero-frequency limit is given through the following relation [28]

$$S(V) = S_0 \int_{-\infty}^{+\infty} d\varepsilon \sum_{m,n} \Big\{ (T_{m,n})^2 \left[ P_m f_L^m (1 - f_L^n) + P_n f_R^n (1 - f_R^m) \right] \\ + (1 - T_{m,n}) T_{m,n} \left[ P_m f_L^m (1 - f_R^n) + P_n f_R^n (1 - f_L^m) \right] \Big\} \tag{8}$$

with $S_0 = 4e^2/h$. In the case of low voltages, the zero-frequency spectral density of shot noise is directly proportional to applied bias $S = PV$, where [48] $P = P_0 T(\varepsilon_F)[1-T(\varepsilon_F)]$ and $P_0 = 4e^3/h$. Another important quantity is the so-called Fano factor defined as $F = S/2e|I|$ which provides information about electron correlations in the system [47]. Here we can distinguish three different cases: sub-Poissonian shot noise with $F < 1$ (electron correlations reduce the level of current fluctuations below unity), Poissonian shot noise with $F = 1$ (there is no correlations among the charge carriers), and super-Poissonian shot noise with $F > 1$ (electron anticorrelations increase the level of current fluctuations above unity). It should be also noted that in the zero-temperature limit, shot noise remains as the only source of electrical noise.

## 3. Results and discussion

To illustrate characteristic features of the transport dependences caused by decoherence, here we consider quantum tunneling process via discrete quantum state in the presence and absence of phonons. This simplified test case is dictated by transparency of analysis to discuss the essential physics of the problem in detail. In our calculations we have used the following set of model parameters (given in eV): $\varepsilon_l = 0$ (the reference LUMO level), $\varepsilon_F = -1$, $\Omega = 1$, $\lambda = 0.5$, $\Gamma_L = \Gamma_R = 0.1$ (the case of symmetric connection to both electrodes, where $\eta = 1/2$), and $\Gamma_D = 0.1$ (as estimated in introduction). Moreover, the temperature of the system is set at $\theta = 300$ K ($\beta^{-1} = 0.025$), while the maximum number of allowed phonon quanta $m_{max} = 8$ is chosen to give fully converged results for all the model parameters. In this work we take into account one-phonon as well as many-phonon processes, where one or even few phonons can be exchanged by electron tunneling through the molecule into the individual act of scattering.

In Fig.2a we plot the linear conductance as a function of Fermi energy. For one discrete energy level, when electron-phonon coupling is neglected, one resonant transmission peak is observed. Phase decoherence decreases the resonant transmission probability ($G_{peak} < G_0$) in comparison with the perfect one-channel transmission ($G_{peak} = G_0$) expected for coherent transport (see the appendix). Besides, the width of the transmission peak as an effect of the molecule-metal connections is also slightly enhanced due to dephasing. For the case of non-zero electron-phonon coupling, the transmission function reveals additional peaks which indicate the opening of channels involving phonons, while the main peak is reduced in height. Positions of the mentioned peaks approximately coincide with polaron energies $\varepsilon_{pol}(m) = \varepsilon_l - \Delta + m\Omega$, where $m$ denotes the $m$th excited state of a polaron (defined as a state of an electron coupled to phonons), while $\Delta = \lambda^2/\Omega$ is the so-called polaron shift. The main peak corresponds to the tunneling process through the polaron ground state $\varepsilon_{pol}(0)$, while additional side peak represents the first excited state $\varepsilon_{pol}(1)$. Obviously, the separation between the conductance peaks is set by the energy of the phonon mode, since $\varepsilon_{pol}(1) - \varepsilon_{pol}(0) = \Omega$. The effect of dephasing is to broaden all the conductance peaks and to reduce their height in comparison with coherent conductance spectrum.

Figure 2b shows the linear shot noise against the Fermi energy. In the absence of phonons, when transport is coherent, the $P$-quantity exhibits two peaks separated by antiresonance and located symmetrically around the position where the conductance peak associated with the single level of the dot is located. The origin of such antiresonance in the noise spectrum is associated with the fact that no noise is generated when the transmission via molecular state is perfect $T=1$ (the same statement is valid for antiresonances in the transmission function $T=0$). Anyway, the effect of dephasing is to broaden the noise peaks and, as its consequence, to destroy the discussed antiresonant state (the spectral dip reaches local minimum of finite value instead of zero). Interestingly, decoherence has no influence on the height of the noise peaks. Including inelastic processes, the $P$-quantity is shifted due to polaron formation by $\Delta$ and exhibits satellite peak at the positive energy side.

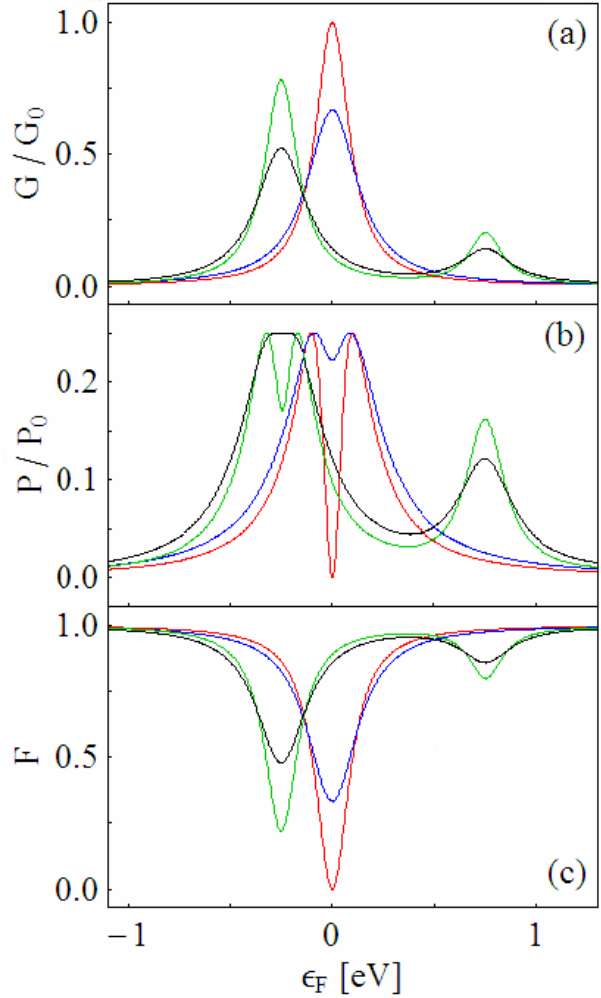

Figure 2: Linear conductance (a), the $P$-quantity associated with the linear noise power (b), and the linear Fano factor (c) as a function of the Fermi energy for four different sets of model parameters: $\Gamma_D = \lambda = 0$ (red lines), $\Gamma_D = 0.1$ and $\lambda = 0$ (blue lines), $\Gamma_D = 0$ and $\lambda = 0.5$ (green lines), $\Gamma_D = 0.1$ and $\lambda = 0.5$ (black lines). The other parameters of the model: $\varepsilon_l = 0$, $\Gamma_L = \Gamma_R = 0.1$, $\Omega = 1$, $\beta^{-1} = 0.025$.

Satellite is represented by only one peak, since the corresponding transmission peak is small ($T \ll 1$) and therefore the noise power is approximately proportional to the conductance (since $T(1-T) \cong T$). Now, decoherence results in joining together two main peaks into a one broad peak of the same height and in broadening the satellite peak of reduced height. It is clear that Fano factor plotted in Fig.2c is a simple reflection of the conductance spectrum (since $F = 1 - T(\varepsilon_F)$), where the conductance peaks are replaced by dips of the Fano factor. Here we can observe the transitions from the Poissonian limit ($F = 1$) to sub-Poissonian region ($F < 1$), while the effect of dephasing is to reduce electron correlations associated with Pauli principle in the vicinity of resonances. Our conclusions retaining to coherent transport are confirmed by the results already known in the literature [41].

All the features of the transmission function discussed earlier are reflected in the nonlinear transport characteristics. In the absence of phonons, one-step current structure occurs, when the electrochemical potential of the left electrode ($\mu_L$) coincide with the LUMO level of the molecular quantum dot ($\varepsilon_l$), as can be seen in Fig.3a. Here we can point out that phase decoherence results in smoothing of the $I-V$ curve (see the appendix). Inclusion of the electron-phonon coupling leads to polaron formation, that is: (i) polaron shift of the main current step, and (ii) additional current step associated with polaron excited state. The first effect results in the reduction of the so-called conductance gap given through the simplified relation: $4(|\varepsilon_F - \varepsilon_l| - \Delta)$. Again, the effect of dephasing is to smooth the $I-V$ dependence. Obviously, the shape of the shot noise curves is similar to that of the currents, as is viewed in Fig.3b.

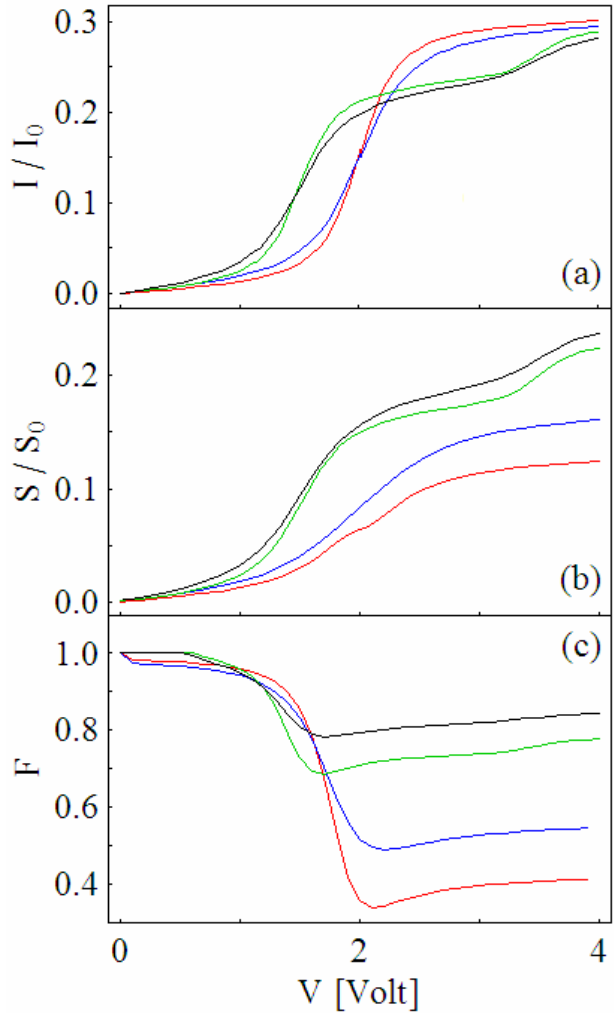

Figure 3: Electrical current (a), nonlinear noise power (b), and the nonlinear Fano factor (c) as a function of bias voltage for four different set of model parameters: $\Gamma_D = \lambda = 0$ (red lines), $\Gamma_D = 0.1$ and $\lambda = 0$ (blue lines), $\Gamma_D = 0$ and $\lambda = 0.5$ (green lines), $\Gamma_D = 0.1$ and $\lambda = 0.5$ (black lines). The other parameters of the model: $\varepsilon_l = 0$, $\varepsilon_F = -1$, $\Gamma_L = \Gamma_R = 0.1$, $\Omega = 1$, $\beta^{-1} = 0.025$.

As was mentioned in the previous section, information about the statistical properties of the electrons is included into Fano factor, which is plotted in Fig.3c. In the absence of phonons, the crossover in the noise power from Poissonian limit ($F=1$) to sub-Poissonian region ($F<1$) is observed right after the step in the $I-V$ function and it means that electrons tunnel in a correlated way for $V>2$. Similarly in the presence of phonons, such transition is observed with polaron shift correction $\Delta$. Since in the sub-Poissonian region the Fano factor reaches higher values for polaronic transport than for elastic transfer, we conclude that electron correlations are partially reduced due to electron-phonon interactions. In analogy, the Fano factor for higher voltages reaches higher values in the case of incoherent transport in comparison with purely coherent transfer, and for that reason we conclude that the inclusion of phase decoherence results in further reduction of electron correlations in the considered system.

## 4. Concluding remarks

In summary, here we have studied the effect of decoherence on elastic and polaronic transport via discrete quantum states. The calculations were performed with the help of nonperturbative computational scheme, based on Green's function theory within the framework of polaron transformation (GFT-PT), where the many-body electron-phonon interaction problem is mapped into a single-electron many-channel scattering problem. In particular, the influence of dephasing and relaxation processes on the shape of the electric current and shot noise curves were discussed in detail. Under the linear transport conditions, the changes due to phase decoherence are associated with: (i) reduction of the height and broadening of the peaks, (ii) destruction of the antiresonance in the noise spectrum, and (iii) partially reduction of electron correlations in the vicinity of resonances. Under nonlinear transport conditions, the changes due to phase decoherence are associated with: (i) smoothing of the transport characteristics, (ii) partially reduction of electron correlations for higher voltages (in the sub-Poissonian region).

It should be also emphasized that the computational scheme presented in this work is based on few drastic assumptions. For example, here we have completely neglected Coulomb interactions between charge carriers, phonon mediated electron-electron interactions or nonequilibrium phonon effects (i.e. distribution function is independent of density of the current). Besides, phase decoherence is treated by means of the model proposed by Büttiker. Anyway, a better model to describe decoherence is to let the broadening due to vibrational coupling at each energy be proportional to the density of states at that energy [36,37]. This requires a self-consistent evaluation of Green function (given by Eq.5) and self-energy defined as $\Sigma_D = \gamma_D G$, where $\gamma_D$ represents the strength of the coupling of the electronic levels to the molecular vibrations.

Inelastic electron tunneling is quite important from the viewpoint of structural stability [49] and the switching possibility of the molecular electronic devices, since this mechanism of conduction may cause chemical bond breaking and chemical rearrangement in the molecular complex. Recently, the negative differential resistance (NDR effect) and hysteresis behavior of the $I-V$ dependence are suggested to be a direct consequence of polaron formation [50]. Moreover, the problem of localized electron-phonon interactions is related to the problem of local heating in current carrying molecular junctions [51,52].

## Appendix

Here we introduce the analytical results for incoherent transport via single quantum state in the absence of phonons. It can be easily shown that the particular components of the total transmission function $T_{tot} = T_{coh} + T_{incoh}$ are expressed as follows

$$T_{coh} = \frac{4\Gamma_L \Gamma_R}{4(\varepsilon - \varepsilon_0)^2 + \Gamma^2}, \tag{A.1}$$

$$T_{incoh} = \frac{4\Gamma_{eff} \Gamma_D}{4(\varepsilon - \varepsilon_0)^2 + \Gamma^2}, \tag{A.2}$$

where $\Gamma_{eff} = \Gamma_L \Gamma_R / (\Gamma_L + \Gamma_R)$, $\Gamma = \Gamma_L + \Gamma_R + \Gamma_D$. Now let us consider the resonance condition ($\varepsilon = \varepsilon_0$): (i) in the limit of $\Gamma_D \to 0$ we have $T_{incoh} = 0$ and $T_{coh} = T_{tot} = 1$ (perfect transmission); (ii) in the case of $\Gamma_L = \Gamma_R = \Gamma_D$ we have $T_{incoh} = 2/9 < T_{coh} = 4/9$ and $T_{tot} = 2/3$ (transmission is suppressed due to the dephasing processes). In the low-temperature limit, the particular contributions to the electrical current flowing through the junction $I_{tot} = I_{coh} + I_{incoh}$ are given through the relations

$$I_{coh} = \frac{2I_0 \Gamma_L \Gamma_R}{\Gamma} \left[ \arctan\left(\frac{2(\varepsilon_0 - \mu_R)}{\Gamma}\right) - \arctan\left(\frac{2(\varepsilon_0 - \mu_L)}{\Gamma}\right) \right], \tag{A.3}$$

$$I_{incoh} = \frac{2I_0 \Gamma_{eff} \Gamma_D}{\Gamma} \left[ \arctan\left(\frac{2(\varepsilon_0 - \mu_R)}{\Gamma}\right) - \arctan\left(\frac{2(\varepsilon_0 - \mu_L)}{\Gamma}\right) \right]. \tag{A.4}$$

The maximal current flow can be calculated by consideration of extremely high voltages ($V \to \infty$), where $I_{coh}(\infty) = 2\pi I_0 \Gamma_L \Gamma_R / \Gamma$ and $I_{incoh}(\infty) = 2\pi I_0 \Gamma_{eff} \Gamma_D / \Gamma$. Similarly we can take into account two different situations: (i) in the limit of $\Gamma_D \to 0$ we have $I_{incoh}(\infty) = 0$ and $I_{coh}(\infty) = I_{tot}(\infty) = 2\pi I_0 \Gamma_{eff}$; (ii) in the case of $\Gamma_L = \Gamma_R = \Gamma_D$ we have $I_{incoh}(\infty) = \pi I_0 \Gamma_R / 3$, $I_{coh}(\infty) = 2\pi I_0 \Gamma_R / 3$ and $I_{tot}(\infty) = \pi I_0 \Gamma_R$. It should be emphasized that if $\Gamma_L = \Gamma_R = 0.1$ eV, the maximal value of the total current is the same in both considered cases $I_{tot}(\infty) \approx 24.3$ μA. Since the height of a step in the $I-V$ curve is directly proportional to the area of the corresponding peak in the transmission spectrum, we conclude that consequences of the dephasing are associated with: (i) reduction of the height of the transmission peak, (ii) broadening of the transmission peak, and (iii) smoothing of the $I-V$ characteristic.